\documentclass[useAMS,usenatbib]{aa}  
\usepackage{graphicx}
\usepackage{txfonts}
\usepackage{natbib}
\bibpunct{(}{)}{;}{a}{}{,} 
\def\deg{\ifmmode^\circ\else$^\circ$\fi}

\def\msun{M$_{\odot}$}
\def\lsun{L$_{\odot}$}

\newcommand{\mum}{\,$\mu$m}

\def\n2h+{N$_2$H$^+$}
\def\c18o{C$^{18}$O}

\def\deg{\ifmmode^\circ\else$^\circ$\fi}

\def\msun{M$_{\odot}$}

\begin{document}
   \title{Spitzer-IRAC GLIMPSE of high mass protostellar objects. I}

   \subtitle{Infrared point sources and nebulae}

   \author{M. S. N. Kumar\inst{1}
          \and J. M. C. Grave\inst{1,2}
		}

   \offprints{M. S. N. Kumar; email:nanda@astro.up.pt}

   \institute{Centro de Astrof\'{i}sica da Universidade do Porto, 
              Rua das Estrelas, 4150-762 Porto, Portugal
	  \and Departamento de Matem\'{a}tica Aplicada da Faculdade de Ci\^{e}ncias da Universidade do Porto, Portugal\\
	  \email{nanda@astro.up.pt; jgrave@astro.up.pt }}

   \date{}

\abstract
{} {To conduct a statistical study of candidate massive protostellar
objects in the 3.6--8.0\mum\, bands of the Spitzer Space Telescope.} 
{The GLIMPSE archive was used to obtain 3.6--8.0\mum\, point source
photometry and images for 381 massive protostellar candidates lying in
the Galactic mid-plane. The colours, magnitudes and spectral indicies
of sources in each of the 381 target fields were analysed and compared
with the predictions of 2D radiative transfer model simulations. } 
{Infrared point sources with intrinsic redenning were found associated
with several massive protostars. Although no discernable embedded
clusters were found in any targets, multiple sources or associations of
redenned young stellar objects were found in many sources indicating
multiplicity at birth. The spectral index ($\alpha$) of these point
sources in 3.6--8.0\mum\, bands display large values of
$\alpha$=2--5. A color-magnitude analog plot was used to identify 79
infrared counterparts to the HMPOs that are bright at 8\mum\,
,centered on millimeter peaks and display $\alpha$ values in excess of
2. Compact nebulae are found in 75\% of the detected sources with
morphologies that can be well described by core-halo, cometary,
shell-like and bipolar geometries similar to those observed in
ultra-compact HII regions.}  {The IRAC band spectral energy
distributions (SED) of the infrared counterparts of massive
protostellar candidates are best described to represent YSOs with a
mass range of 8--20\msun in their Class I evolutionary stages when
compared with 2D radiative transfer models. They also suggest that the
high $\alpha$ values represent reprocessed star/star+disk emission
that is arising in the dense envelopes. Thus we are witnessing the
luminous envelopes around the protostars rather than their
photospheres or disks. We argue that the compact infrared nebulae
likely reflect the underlying physical structure of the dense cores
and are found to imitate the morphologies of known UCHII regions. The
observations are consistent with a scenario where massive protostars
have formed inside dense cores and continue to accrete matter. Our
results favour models of continuuing accretion involving both
molecular and ionised accretion components to build the most massive
stars rather than purely molecular rapid accretion flows.}


   \maketitle
%

\section{Introduction}

The Spitzer Space Telescope with its IRAC camera is bringing in a
revolution to our understanding of the star formation process by
opening a great depth in sensitivity and mapping speed at the
3.6--8.0\mum\, region of the infrared spectrum. The Galactic mid-plane
has been mapped in these bands through the Science Legacy Program
GLIMPSE. Much of the massive star formation in the galaxy occurs in
its mid-plane; the availability of the point source photometry catalog
and the {\em image cutouts} facility from the GLIMPSE program thus
gives us a chance to study the statistical properties of candidate
massive protostars.

In more than a decade long effort to study the early stages of massive
star formation, four major surveys to search for massive protostellar
candidates have been completed covering both the northern and southern
hemispheres of the sky. These surveys are described in \citet{mol96},
\citet{sri02}, \citet{fon02} and \citet{faun04}. Hereafter we will
refer to them as Sri02, Mol96, Fon02 and Faun04.  Follow-up
studies of many of these targets to study the dust continuum
properties have also been made \citep{beu02,beltran06,mol00}.  All these
surveys are based on a selection criteria using constraints on
far-infrared (FIR) colours and choosing additional sign-posts of star
formation, in an effort to identify phases of massive star formation
prior to the onset of an ultra-compact HII (UCHII) region. The surveys
are therefore, all based on FIR and millimeter observations that can
probe into great depths of column density usually found in the dense
cores that form massive stars. However, the large distances at which
massive star forming regions are typically found and the moderate beam
sizes of FIR and millimeter telescopes used for these surveys have
limited the spatial size scales of observations to typically 0.1~pc or
larger. Dense cores with such physical dimensions may host a single
high mass protostellar object (HMPO) or a cluster of embedded
stars. For example, using near-infrared (NIR) 2MASS data,
\citet{kumar06} showed that 25\% of the HMPOs from
Mol96 and Sri02 surveys are associated with embedded clusters. The
Spitzer-IRAC 3.6--8.0\mum\, observations from the GLIMPSE survey
provides an extra edge over the previous FIR and mm observations to
probe into smaller spatial dimensions. At a comparable spatial
resolution to the 2MASS data, the 8\mum\, GLIMPSE data can penetrate
into extinction depths approximately four times better than the 2MASS
K band data assuming sensitivity limits of 14 and 12 magnitude from
2.2 and 8\mum\ and using the measurement that A$_{8\mu m}$/A$_{K}$=0.5
\citep{flaherty07}.  The full-width half maximum of the Spitzer-IRAC
point spread function is $\sim$2.4\arcsec which at a distance of
5\,kpc will trace a projected size of $\sim$10,000\,AU. This is
typically the size of an extended envelope or a toroid around a HMPO.

In this article (Paper I), we present a statistical analysis of the
mid-infrared (MIR) properties of about 381 HMPO candidates, based on
the publicly available GLIMPSE database. We aim to identify the
infrared counterparts of the millimeter cores, evaluate their colour
properties and search for embedded clusters associated with the HMPO
targets. The emission from disk and envelope regions around young
stellar objects (YSO) are known to deviate significantly from that of
a normal star in the 3.6--8.0\mum\, bands allowing the YSO
classification into various evolutionary stages. While the millimeter
observations are more useful to estimate the mass of the dense cores
in which HMPOs are born, the GLIMPSE data will enable us to constrain
the evolutionary stages and mass ranges of any point sources
associated with HMPOs. In Paper II \citep{grave07} we will present an
analysis of various physical parameters of well identified individual
sources derived from modelling the infrared-millimeter spectral energy
distribution using radiative transfer models.
    
\section{Data selection and analysis}

The sample of objects are derived from the four surveys of HMPOs as
listed in Table.\,1. An estimate of the FIR luminosity and distance is
available for most of the targets within the usual uncertainity
factors. Distance estimate is missing for about 60 targets from
Mol96 and Fon02 are not available from the original papers of which
about 20 targets were detected in GLIMPSE. For such targets we have
assumed a fiducial mean distance of 5kpc. As evident from Table.\,1,
data in the GLIMPSE catalog was available for $\sim$80\% of all
the targets. We used both the GLIMPSE point source catalog (highly
reliable) and the GLIMPSE point source archive (less reliable and more
complete) to search for available data. Photometric data in the four
IRAC bands at 3.6, 4.5, 5.6 and 8.0 $\mu$m were obtained centered on
each target.

For the purpose of cluster detection, point source data was extracted
in a 600\arcsec\, diameter circle centered on each target and a
detection in any one of the four bands was considered as a source. For
the purpose of identifying infrared point sources and analysing their
spectral properties, data was extracted from a region of 100\arcsec\,
radius centered on each target with the constraint that the source be
detected in at least 2 of the four bands with a minimum of 3$\sigma$
detection in each of the four bands. There is one important caveat in
using the point source photometry catalog. Several HMPO candidates are
very bright in the IRAC bands (particularly in the 8$\mu$m band)
resulting in a saturated signal on the images and null data in the
point source catalog. Such objects may indeed be the best counterparts
of the HMPO targets in the list but will not be found in our
analysis. However, such sources can be seen on the actual images where
a saturated bright red-star and/or a compact nebulae can be found
associated with FIR and millimeter peaks.

\begin{table} 
  \caption[]{Target Statistics}
  \begin{tabular}{@{}lrrrr@{}}
  \hline
  \hline
Name of the & Number of & Number found & Number of\\
Survey   & targets & in GLIMPSE & Nebulae\\
\hline
\citet{mol96} & 161 & 79 & 59\\
\citet{sri02} & 69 & 48 & 48\\
\citet{fon02} & 133 & 110 & 76\\
\citet{faun04} & 146 & 144 & 105\\
\hline
Overall Statistics & 509 & 381 & 288\\
\hline
\end{tabular}
\end{table}

\section{Clustering around HMPO targets}

The point source data extracted in a 600\arcsec\, circle around each
target was Nyquist binned with bin sizes of 60\arcsec\, and
120\arcsec\, and contours of 2$\sigma$ and above were plotted to
identify any clustering around each target. Data was available for 381
out of 500 targets derived from all the four surveys for which star
count density enhancements were checked. No significant clusters were
detected around any of the targets although weak associations emerged
when point sources only in the 5.8\mum\, band were considered. Given
the non-consistency of such weak detections depending upon the
photometric band used to obtain the point sources, we conclude a null
result for cluster detections for the targets surveyed. In contrast,
using a similar technique with 2MASS point source catalog,
\citet{kumar06} were able to detect 54 clusters from the Mol96 and
Sri02 samples, for all the targets that {\em fell away} from the
galactic mid-plane.\citet{kumar06} survey could not find any clusters
in the Galactic mid-plane owing to the high extinctions in this
region. While the GLIMPSE survey could easily probe in to the Galactic
mid-plane and found several infrared counterparts and nebulae which
will be discussed in following sections, the lack of cluster detection
is due to a reduced sensitivity to lower mass stars and an increased
contamination of background and foreground star counts. The GLIMPSE
survey data is sensitive only to 2-4\msun\, pre-mainsequence stars at
a distance of 3kpc owing to the short exposure times of 2 seconds in
each photometric band. Embedded clusters around massive stars are
mostly composed of lower mass components which will not be detected at
this sensitivity. Therefore the null result here only implies the
absence of major embedded clusters with massive members and/or
ensembles of several hundred members.

\begin{figure}
   \centering \includegraphics[width=9cm]{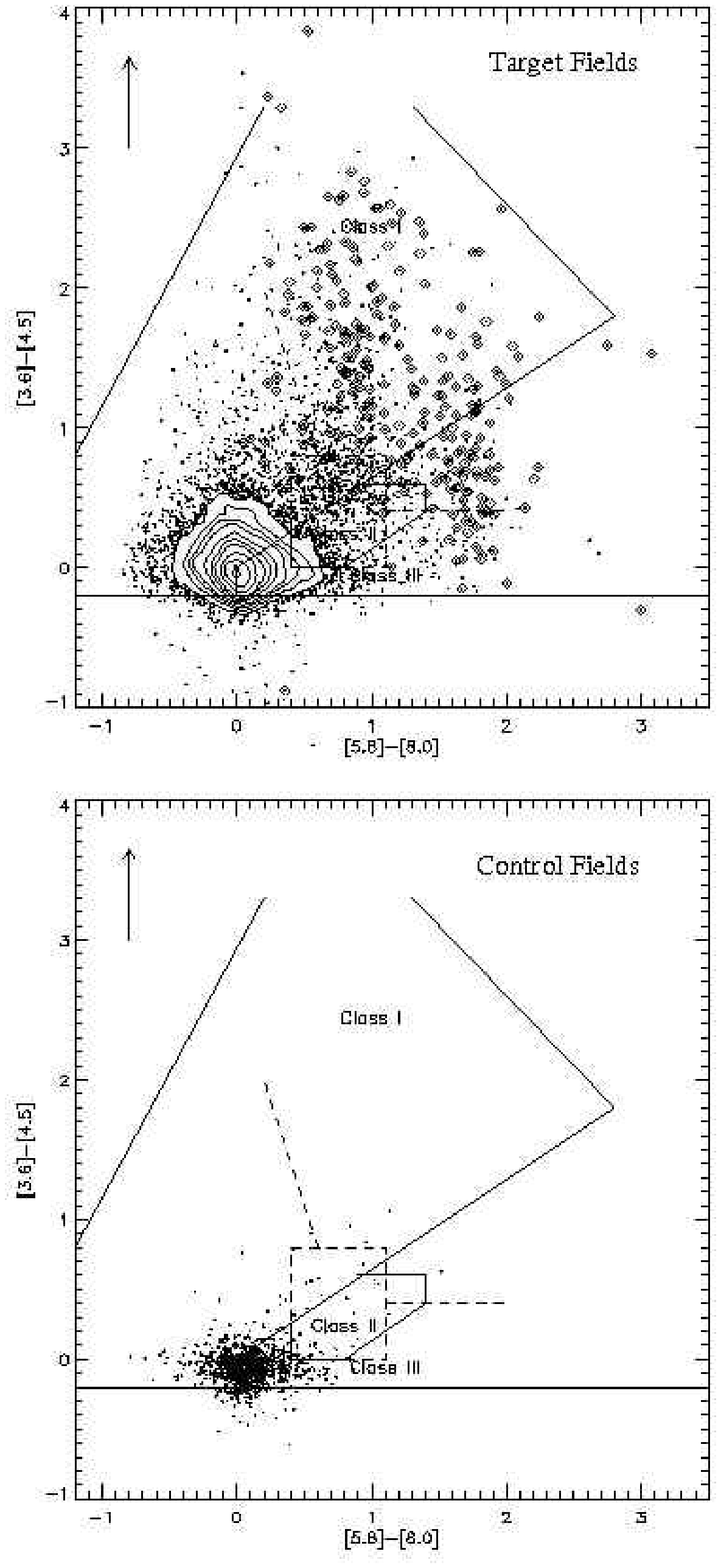}
   
   \caption{Top panel: Color-Color diagram of IRAC point sources
   associated with candidate massive protostars. Contours are used to
   better display the density of points at it highest values near
   (0,0). The box enclosed by dotted lines show the ``disk domain''
   and the dotted lines show the domain of embedded young stellar
   objects. The arrow shows a reddenning vector of length A$_K$=5mag
   according to \citet{indeb05}. The solid lines mark the regions
   occupied by various evolutionary stage YSOs according to the models
   of \citet{robitaille06}. The class I, II and III zones are
   marked. Bottom panel: Same as above but for the data from 40
   control fields placed 0.5\deg adjacent to randomly selected target
   fields from the four surveys. }

  \end{figure}

\section{Point sources associated with HMPO targets}

Although no prominent clusters were detected in the target fields, a
large number of point sources were detected in each target field of
100\arcsec\, radius with good detections in all four bands. The
100\arcsec\, limit on the radius was chosen because the dust continuum
emission, mapped in the millimeter wavebands generally display
emission restricted to this limit indicating that the cores/clumps are
fairly compact around the IRAS sources. Thus, any MIR point sources
within these limits may be associated with such dense cores/clumps
forming the massive protostars. It is important to note that all the
four surveys have selected HMPO candidates as ``isolated luminous IRAS
point sources'' meaning that the FIR emission is localized and point
like. Also, almost all these targets are situated away from major HII
regions and in some cases at the boundaries or outskirts of well known
HII regions. Therefore, while some of the targets may be contaminated
with lower mass younger protostars, most infrared point sources found
in the target fields are likely candidates of intermediate or massive
protostars. In the following we will use the color-color and
color-magnitude analysis to converge on candidate massive protostars.

\subsection{Colors and Spectral Indicies}

Figure.~1 (top panel) shows a [3.6-4.5] vs [5.6-8.0] color-color
diagram for all the sources detected from all target lists. The main
concentration of points at (0,0)(shown using contours) represent
photospheres and higher values on both x and y axis represent YSOs
considered to be at different evolutionary stages. The solid lines
delineate the regions typical of sources at different evolutionary
stages such as I, II, and III based on the 2D radiative transfer model
data of \citet{robitaille06}(hereafter RWIWD06). It can be seen that
there are sources occupying regions representative of all the three
classes namely I, II, III with a significant fraction inside the Class
I and II zones. The diamond symbols encircling the points represent
candidate massive protostars based on a selection criteria that is
described in Sec.4.2.

The bottom panel in Fig.~1 shows the colors of point sources extracted
from a total of 40 control fields; with 10 random fields located 0.5
degrees adjacent to the target fields in {\em each} of the survey
lists. The control fields were similar in area to the target fields
with a radius of 100\arcsec. It can be noted that the sources in these
control fields represent mostly pure photospheres except for some
contamination in the Class II and Class III zones. The Class I zone is
empty in the control field plot. The contamination in the Class II and
III zones in the control fields could be because some of the HMPO
targets are located in the vicinity of larger HII regions which can
contain infrared excess sources. However, the highest infrared excess
sources such as Class I objects are only associated with the target
fields that are known to encompass dense cores with an embedded
luminous FIR source.

Although the color-color diagram is useful to identify infrared excess
emission of the observed targets, a quantitative measure is to compute
the spectral index $\alpha$ of these sources in the observed bands. By
using a simple least squares linear fit to the observed spectral
energy distribution, we computed the spectral index $\alpha$ in the
four IRAC bands for each point source. Figure.~2 shows a histogram of
the relative distribution of $\alpha$ values for the point sources
associated with the massive protostellar targets. A similar
distribution obtained for Spitzer-IRAC sources in the IC348 cluster
and for the YSOs in the Orion A cloud are shown for comparision. For
example, in nearby embedded cluster IC348 \citep{lada06} (3Myr old low
mass cluster), the $\alpha$ values are less than 2 because the sources
are less embedded and relatively more evolved at Class II or III
stages. In the Orion A cloud (1 Myr old and more massive cluster), the
YSOs (note that the photospheres are removed for clarity) show
$\alpha$ relatively higher compared to IC348 sources. The YSOs in
Orion are more embedded and are relatively younger at Class I and II
stages in Orion A. The point sources associated with the massive
protostellar candidates show $\alpha$ values between 2 and 5 and some
even at 6. Such high $\alpha$ sources have been observed in a wide
field study of the massive star forming region DR21/W75N where the
high $\alpha$ sources were found to be well correlated with the
location of HMPOs and/or UCHII regions \citep{kumar07}. Thus the
$\alpha$ values serve as a good indicators of the density of the
surrounding medium in which the sources are embedded. Deeply embedded
Class I sources are also known to produce high $\alpha$ values which
has been observed in the Serpens and Orion regions
\citep{megeath07}. We shall discuss the relevance of high $\alpha$
values for this particular case in Sec.6.

It is evident from the results above that the spectral index is only
representative of the slope of the spectral energy distribution which
can be produced by point sources of different luminosities. Infact a
low mass young stellar object viewed through a column of high optical
depth can produce a large spectral index if detected with sufficient
sensitivity.  The interest in this study is to identify massive
protostellar candidates and for a target situated at a fixed distance
and viewed through a fixed extinction, it is the brightness or
luminosity of the source that identifies the massive
objects. Therefore a color-magnitude diagram is needed to isolate
sources with high luminosity as well as large infrared excess.

  \begin{figure} \centering \includegraphics[width=8cm]{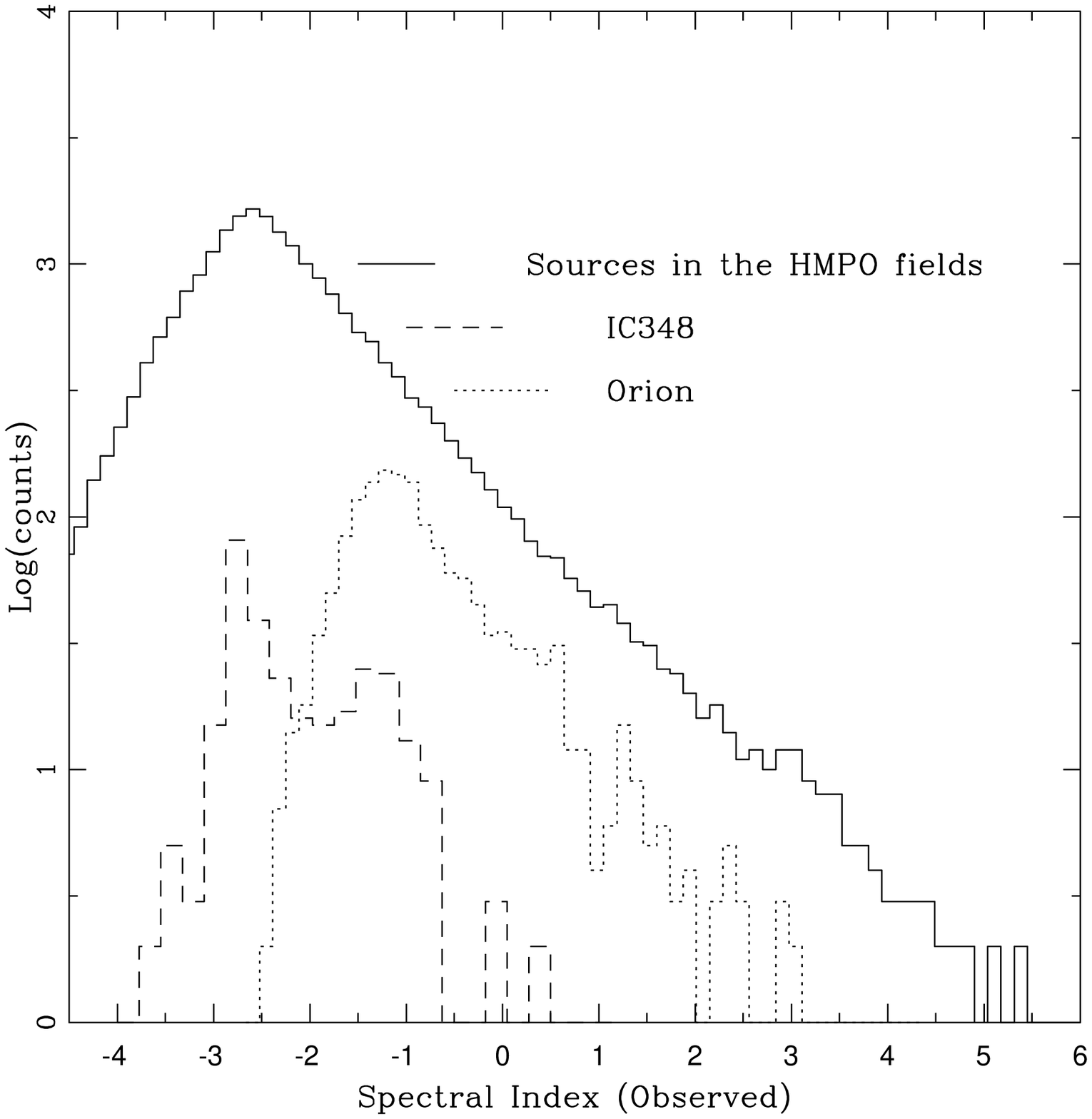}
   
   \caption{Histogram of the observed spectral indicies ($\alpha$) for
   sources associated with candidate massive protostellar
   candidates. The dotted curve shows a similar histogram for the IRAC
   YSOs in Orion A cloud \citep{megeath07} and the dashed curve shows
   the distribution of sources in IC348 \citep{lada06}.}

  \end{figure}

   \begin{figure}
   \centering
   \includegraphics[width=9cm]{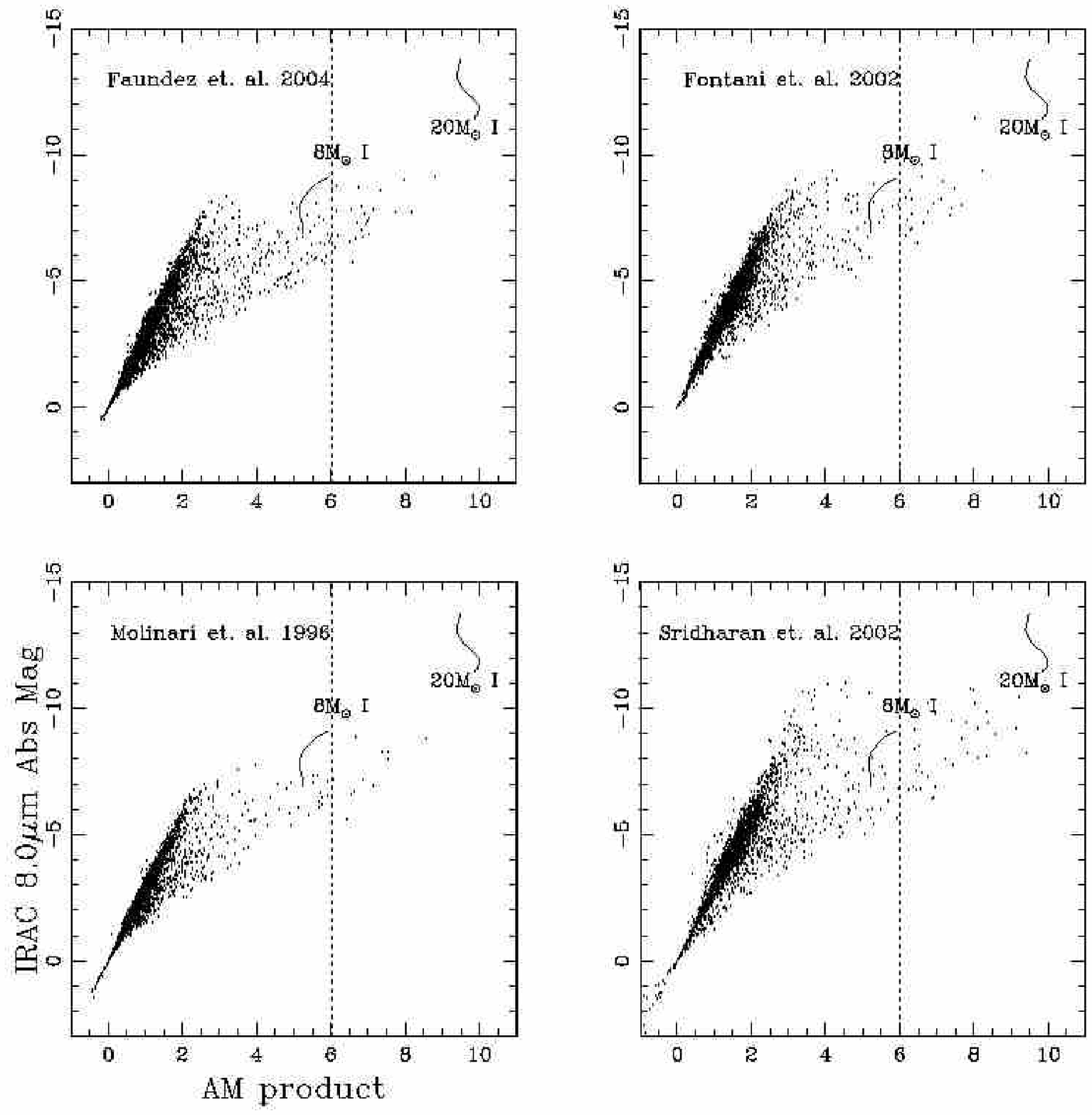}
   
   \caption{The AM product plotted versus the [8.0]$\mu$m absolute
   magnitude for the sources detected from the four surveys of massive
   protostellar candidates. The solid curves represent 20\msun and
   8\msun class I object model curves for all inclinations computed by
   RWIWD06. }

  \end{figure}

\subsection{Spectral index vs Magnitude diagrams}

The sources in our target list are at different distances and a
comparision of colors and magnitudes is possible only by correcting
for distance and line-of-sight extinction. The distances to individual
targets and luminosities were obtained from the respective survey
papers listed in Table.1. However, the line-of-sight extinction to
each source is not easily available and its estimation can vary
depending upon the method and wavelength used to make the extinction
measurement . We therefore applied only the distance correction and
converted the photometric magnitudes to absolute magnitudes for each
source using the appropriate distance. Since the extinction correction
is not made and the IRAC [5.8-8.0] colour shows no difference for a
reddenned object (note that the reddenning vector is vertical) some
luminous reddenned background giants can contaminate the
sample. However, these contaminations can reduce when objects are
selected through the AM product criteria described below since the
spectral index as a whole enter into the AM product.

Next, we found that a simple colour magnitude diagram is confusing
because of a large scatter of points around the region occupied by
main-sequence stars. For example the scatter due to observational
errors and different line-of-sight extinction shows a significant
spread at (0,0) in the color-color diagram shown in Fig.~1 which
translates to a spread of similar width on a color-magnitude
diagram. To reduce this scatter effect, and to separate more
effectively the luminous members, we define a product called the ``AM
product'' (Alpha-Magnitude product), meaning the product of the
spectral index $\alpha$ and absolute magntidue for a source. For
convenience of plotting, we define AM = -M$_{8\mu m}\times$
($\alpha$+6)/10 where M$_{8\mu m}$ is the 8.0\mum\, absolute magnitude
of the source and $\alpha$ is the observed spectral index. The
constants 6 and 10 in the above equation were chosen arbitrarily to
separate the high $\alpha$ sources from the rest effectively. For
example a source of M$_{8\mu m}$=-5 with $\alpha$ values of 0 and 3
will have AM product of 3 and 4.5. Recall the necessity of computing
AM product to obtain luminous sources along with their high colours
rather than separating due to colours alone as explained early in this
section. Figure.~3 shows AM vs 8.0\mum\, absolute magnitude plots for
the sources associated with targets from the four different
surveys. Two model curves for a 8\msun\, and a 20\msun\, Class I
objects from RWIWD06 are shown in this figure for comparision with the
observed points. The model curves represent the locus of points with a
particular mass and for all inclination angles.

It can be seen from Fig.~3 that while most points are concentrated
around the dense slanting branch representative of photospheres, a
significant fraction occupy the zones representative of young stellar
objects. The area encompassed by the curves representing Class I
sources of 8\msun and 20\msun are well separated from the rest of the
points, suggesting the presence of massive protostars. All sources
with AM product higher than 6 delineated by the vertical line in
Fig.~3 are considered as candidate infrared counterparts to HMPO
targets.  A total of 79 point sources with photometry available in all
four IRAC bands could be classified as HMPO targets. The number of
HMPOs detected in Mol96, Sri02, Fon02 and Faun04 are 11, 27, 23 and 25
respectively. This count does not include the brightest counterparts
which are saturated in the IRAC bands. More rigorous isolation of HMPO
sources using measurements at other wavelengths and constructing SEDs
over larger wavelength range will be discussed in Paper II. Figure.~4
shows a spatial plot of the high AM product sources with respect to
the IRAS PSC central coordinates for each target field. It can be seen
that while there is some spread of these sources within an 100\arcsec
box, most sources with high AM values are centered on the IRAS peaks
in the target fields. The associated millimeter dust continuum peaks
also shows some scatter with respect to IRAS centers (e.g. see
\citet{beu02}). The observed scatter is consistent with the spread of
the centers of the 1.2mm peaks from the respective IRAS
positions. Thus, the correlation between the IRAS centers and the
IRAC-point sources with high AM products suggest that we are indeed
observing some of the MIR counterparts of the candidate massive
protostars.

In summary, most of the high AM sources in the Class I region of
Fig.~1 are very good candidates, although the high AM sources with
[5.8-8.0]$<$0.4 colours could also be reddenned photospheres. Some of
the candidates are likely evolved protostars and some can imitate red
colours due to nebular contamination; therefore a complete SED
analysis will be utilised to remove such confusions in Paper II.

   \begin{figure}
   \centering
   \includegraphics[width=8cm]{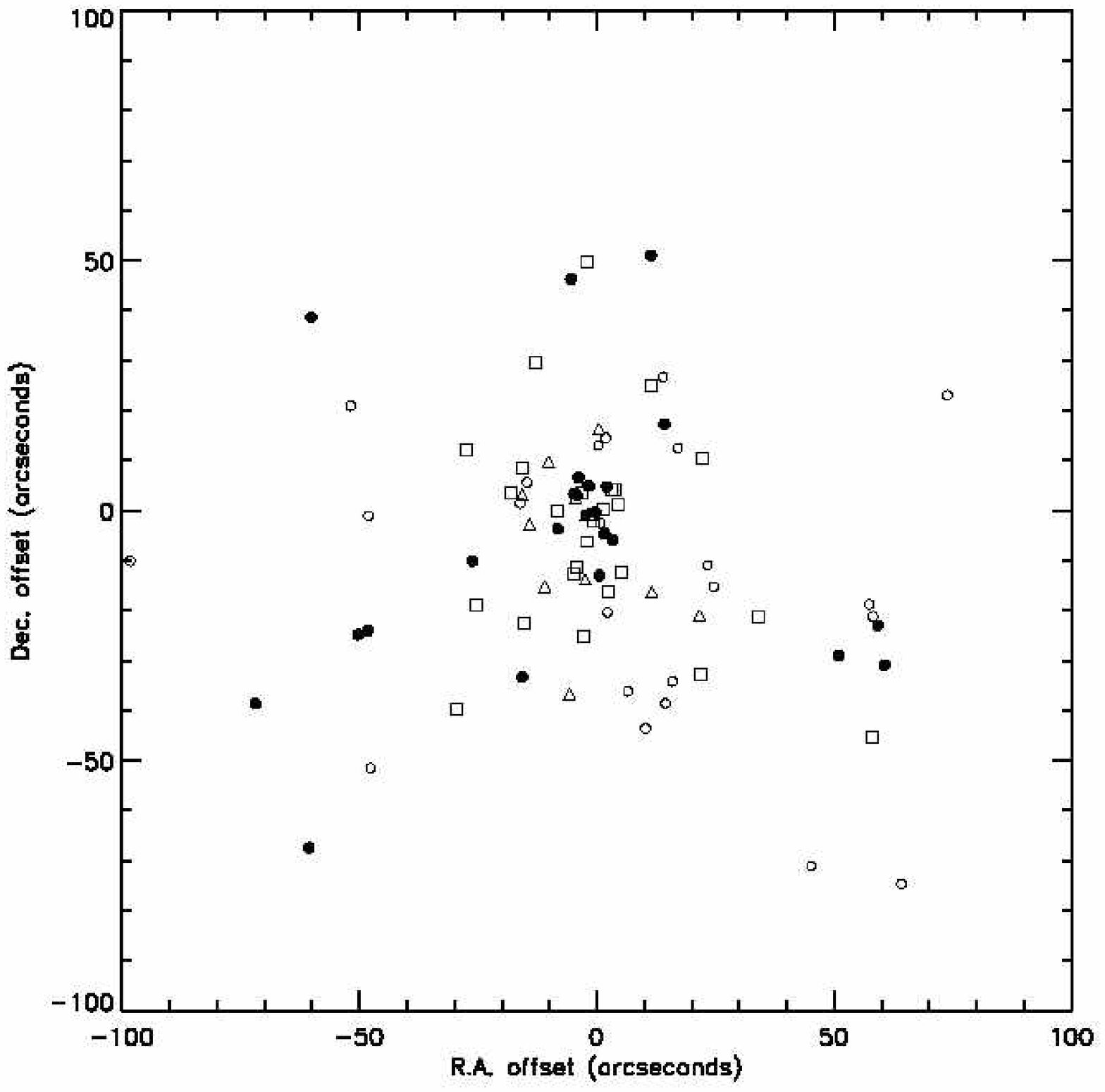}
   
   \caption{Spatial distribution of high AM sources with respect to
   the central coordinates of the associated IRAS sources. Targets
   from all the four surveys namely Mol96 (triangles), Sri02
   (squares), Fon02 (open circles) and Faun04 (filled circles) are
   plotted. Note that bright saturated targets which comprise a
   majority of high AM sources do not appear on this plot.  }

  \end{figure}

\onlfig{5}{ 
  \begin{figure*} \includegraphics[width=19cm]{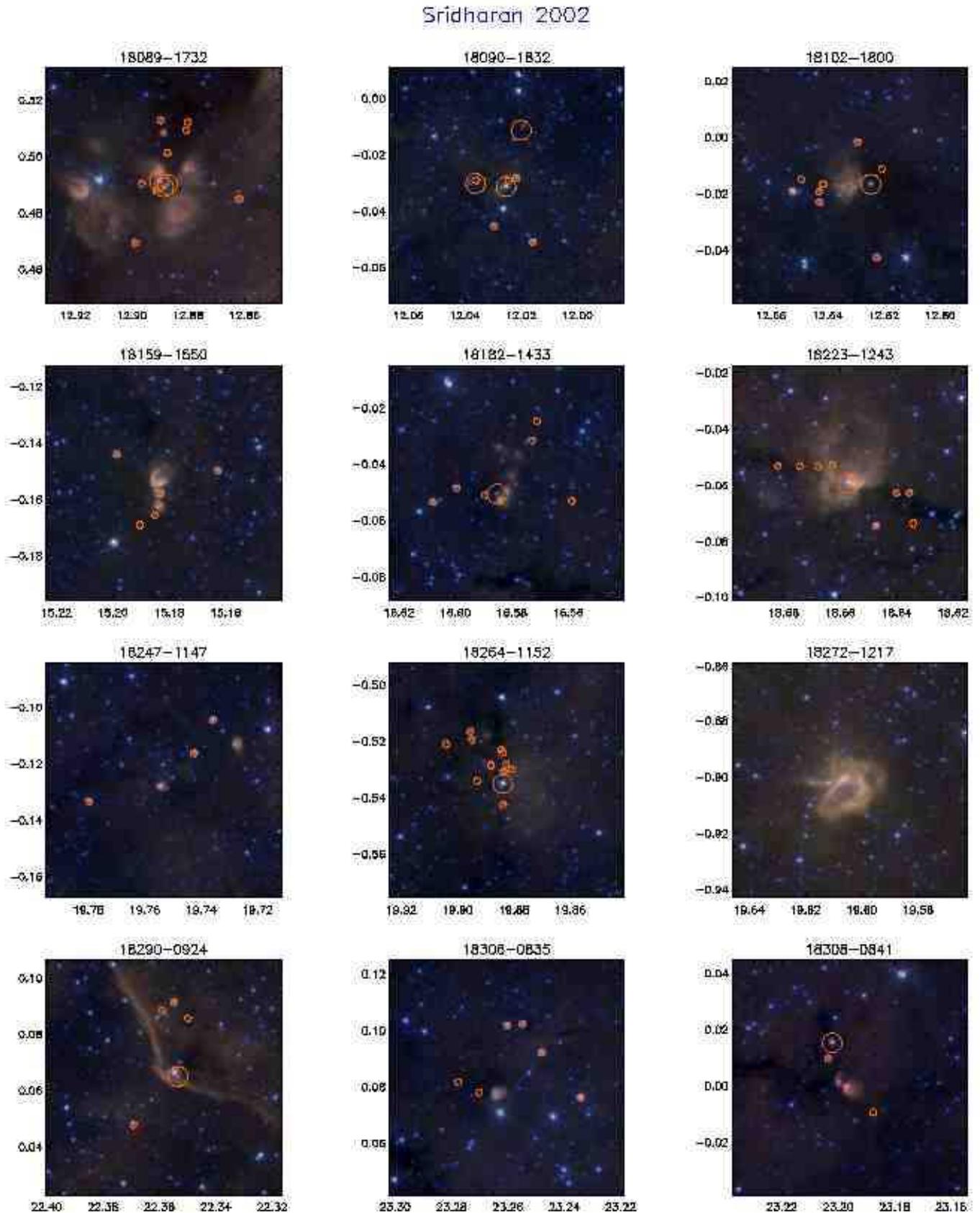}
  \caption{Three color composites of the targets found from
  \citet{sri02} survey. The 4.5\mum\, 5.8\mum\, and 8\mum\, channels
  are coded blue, green and red respectively. Large circles represent
  sources with AM$>$6 and small circles show sources with 4$<$AM$<$6.}
  \end{figure*}

\addtocounter{figure}{-1}
  \begin{figure*}
   \includegraphics[width=19cm]{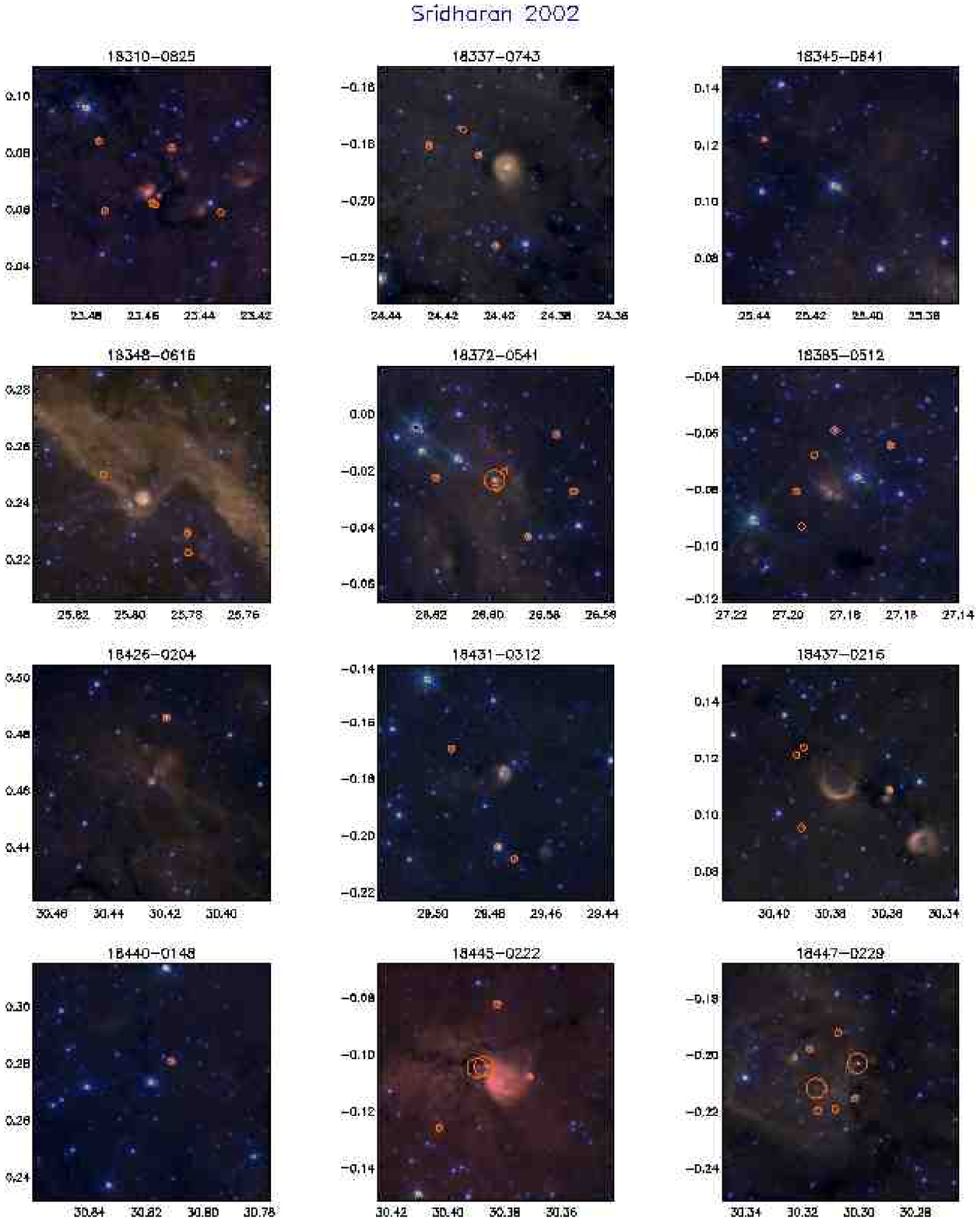}
   \caption{Continued}
  \end{figure*}
\addtocounter{figure}{-1}
  \begin{figure*}
   \includegraphics[width=19cm]{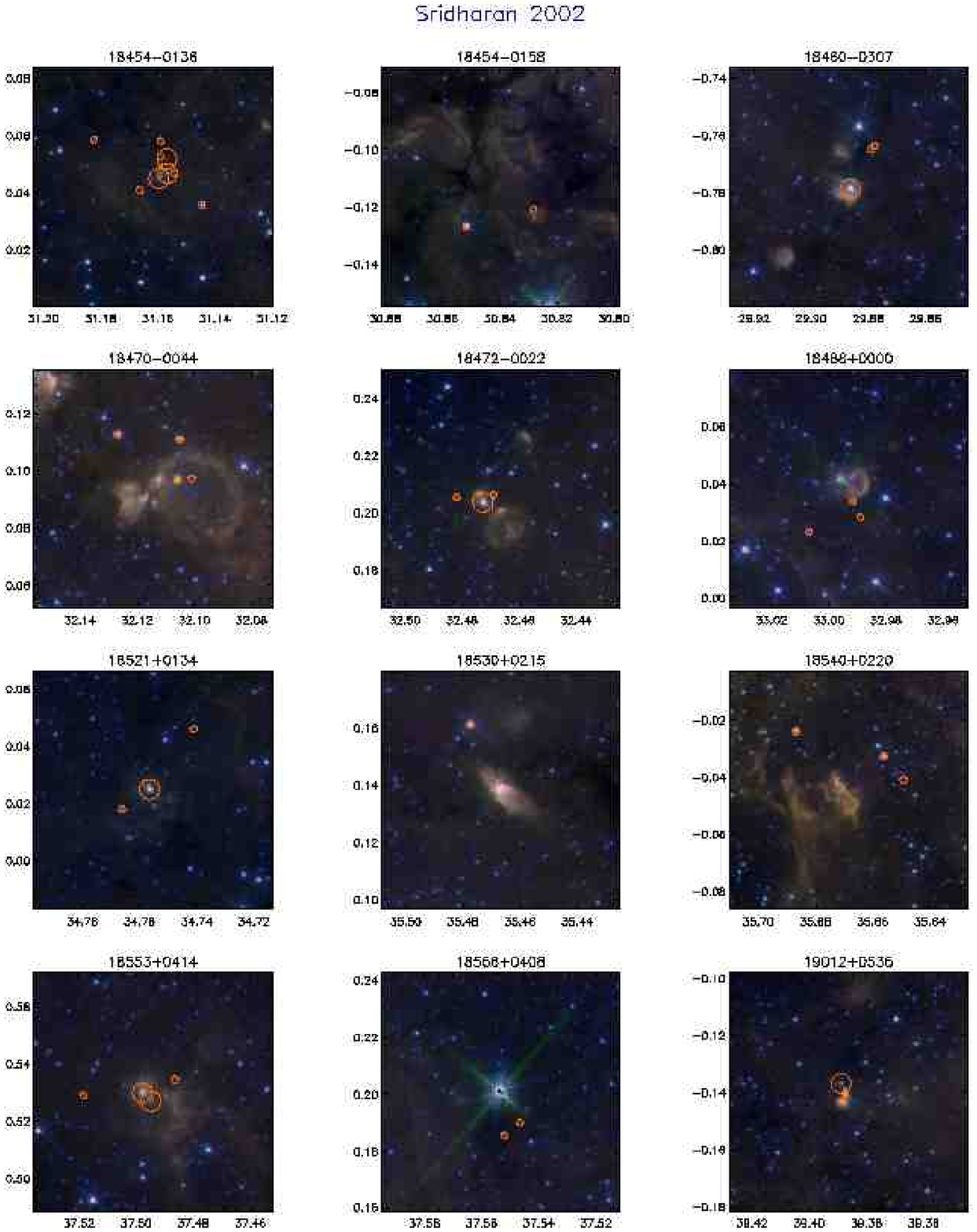}
   \caption{Continued}
  \end{figure*}
\addtocounter{figure}{-1}
 \begin{figure*}
   \includegraphics[width=19cm]{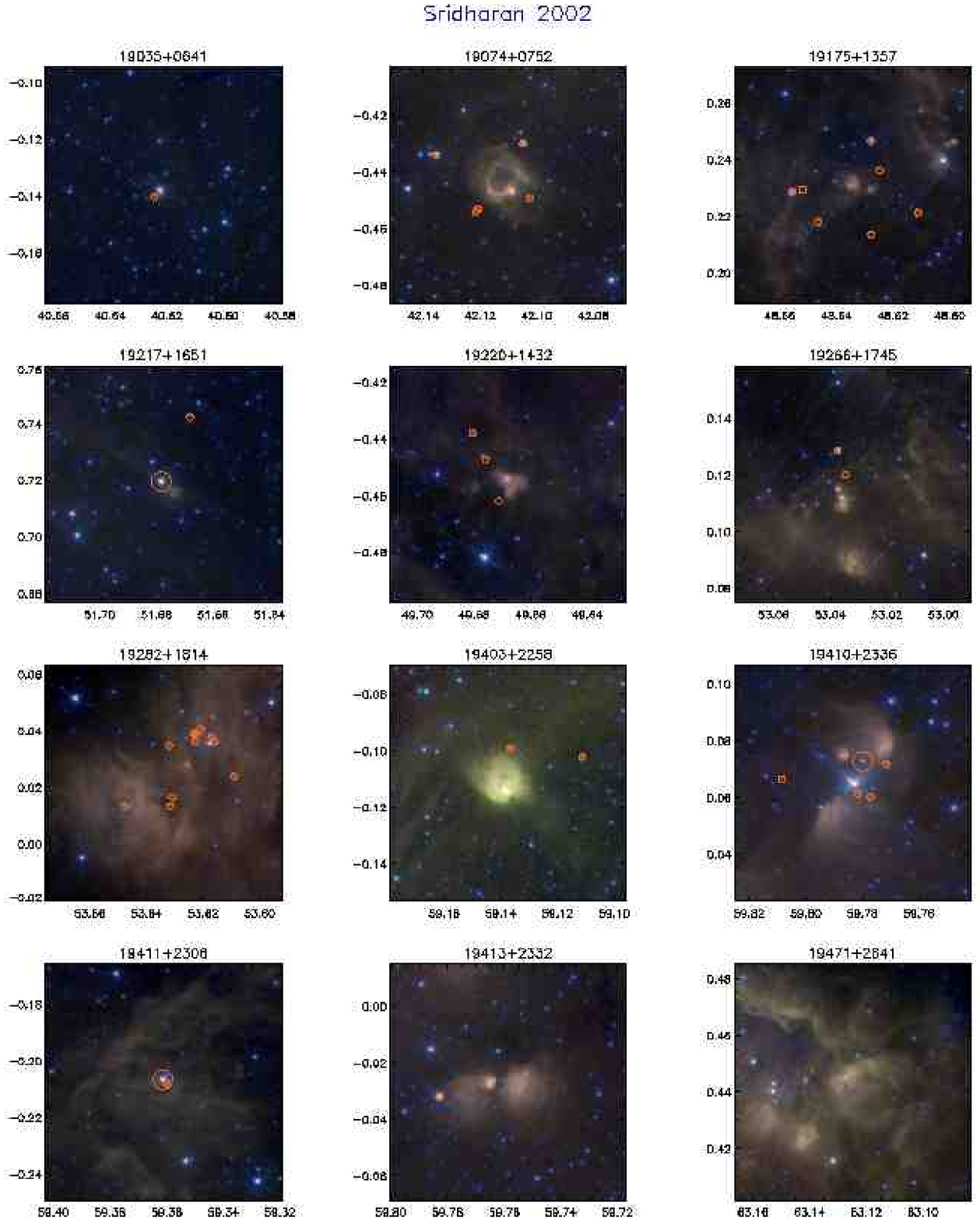}
   \caption{Continued}
  \end{figure*}}

   \begin{figure}
   \centering
   \includegraphics[width=7cm]{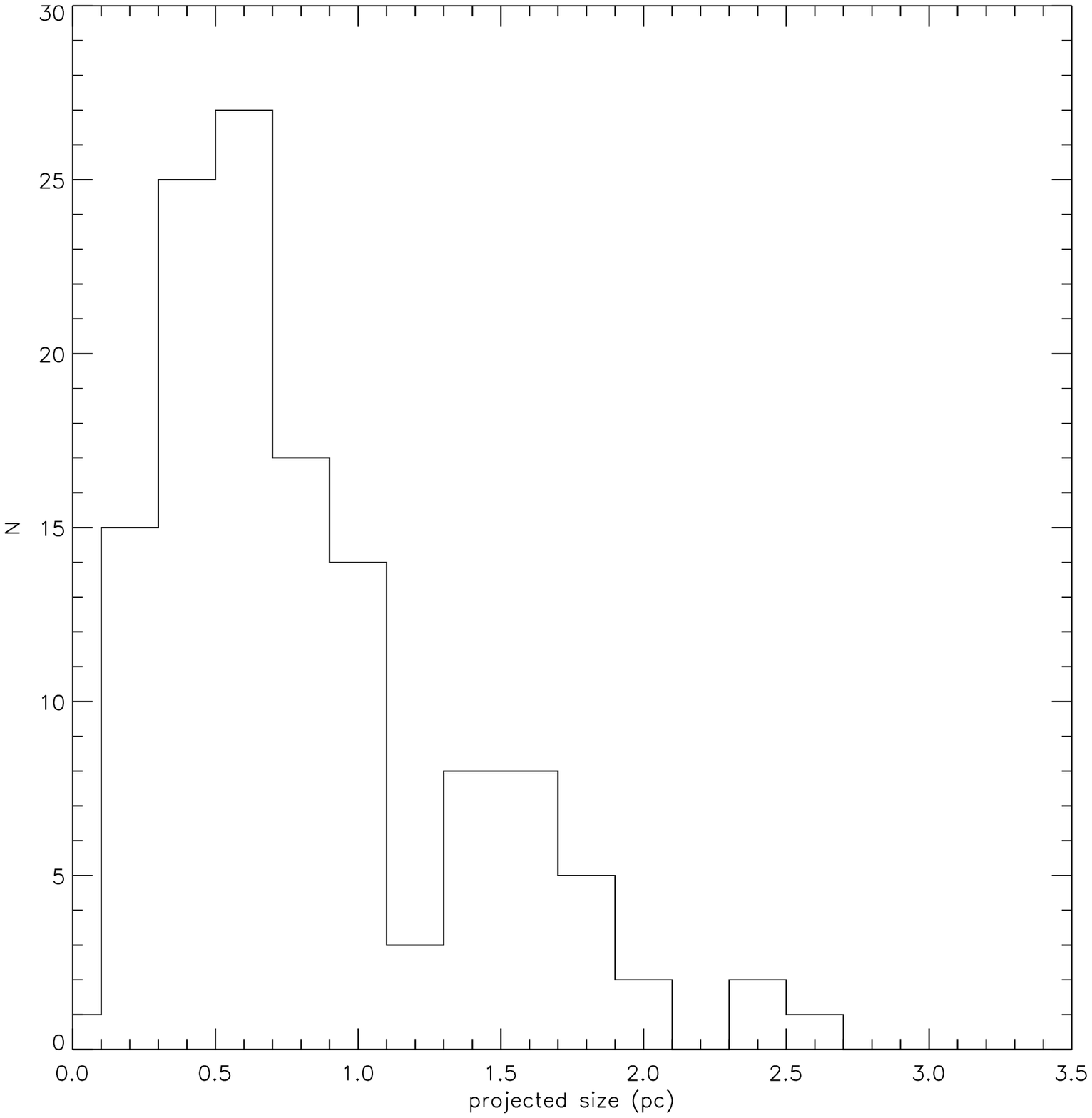}
   
   \caption{Histogram of projected sizes of the infrared nebulae. }

  \end{figure}

\section{Compact Nebulae}

As mentioned in Section.~2, many infrared counterparts to HMPOs can be
very bright in the IRAC bands saturating the detectors, resulting in
null data in the photometry catalog. Also, interesting nebulae around
HMPOs can not be analysed using point source photometry catalog. We
therefore used the { \em image cutouts} facility available on the IRSA
webpage to retrieve IRAC images of size 300\arcsec\, centered on each
target. These images reveal compact nebulae (10\arcsec-60\arcsec
angular sizes) around several targets. Detection statistics of these
nebulae are listed in Table.~1. The nebulae are found to be brightest
in the 8\mum\, band and mostly invisible in the 3.6\mum\, IRAC
band. We have used the 4.5, 5.8 and 8.0 \mum\, band images coded as
blue, green and red respectively to generate composite false color
images for the observed nebulae. The false color images for
\citet{sri02} targets are shown in Fig.~5 (available
online)\footnote{The color composites for the Mol96, Fon02 and Faun04
surveys can be accessed at http://www.astro.up.pt/~nanda/hmpo/}. In
these figures large circlular symbols represent sources with AM values
greater than 6 and small circles represent sources with AM values
between 4 and 6. In some cases, although the circular symbols are
absent, a bright saturated star can be found at the center of the
field, which corresponds to the massive protostars. In Paper II, these
sources will be modelled using data from 2MASS and mm bands as well.

Among the four target samples, Sri02 sample is supposed to represent
the youngest HMPO candidates as that sample used a criteria to select
targets without any significant radio continuum emission. Mol96 and
Fon02 samples are known to have a mixture of source with and without
significant radio continuum emission while Faun04 list contains
several well-known UCHII regions. Although most of the nebulae in our
figures are restricted to the extracted 300\arcsec region, the nebulae
in Sri02 which represents the youngest sample are relatively more
compact.  A careful examination of Fig.~5 will reveal that the
nebulae associated with HMPO targets repeatedly display cometary
(e.g.: I18437-0216), core-halo (e.g: I18337-0743, I19403+2258), shell
like (e.g.: I19198+1423), gaussian (e.g: I19074+0752) and bipolar
(e.g.:I18530+0215, I19213+1723) morphologies. A rough size of these
nebulae was measured on each 8\mum\, image using the ruler option on
the SAOIMAGE DS9 image display widget. Some images display only a
single well defined nebula whereas some others show a group of compact
nebulae. In such cases where multiple components were found, we
measured the mean size of the smaller components and also the upper
limit of the overall size inside which the small components are
embedded. A histogram of the angular scales converted to projected
distances of these nebulae are shown in Fig.~6. The sizes are
typically in the range 0.1-1.0~pc with a mean value of 0.5~pc. Even
the most extended nebulae projects to a maximum of 2.5~pc. Therefore
the dimensions of the infrared nebulae are similar to or smaller than
the size of dense cores as traced by the 1.3mm continuum maps.  UCHII
regions display similar morphologies on high resolution radio
continuum images \citep{kurtz94} and bipolar shapes are representative
of outflows. The 8.0\mum\, band is dominated by PAH emission which is
known to be a tracer of radiation temperature. Ionising radiation from
young massive stars, that may not yet be strong enough to produce a
significant ionised region may therefore be traced by these infrared
nebulae. Indeed a recent investigation has shown that the underlying
structure of the ISM in such nebulae can possibly be inferred using
the morphology of the nebulae at various density regimes and ionizing
fluxes \citep{heitsch07}. Therefore, the morphology of the nebulae
found here may well indicate the morphology in which ionising
radiation is escaping from the underlying set up of physical
structures close to the star. Recently \citet{churchwell06} have used
the GLIMPSE images to identify bubbles around OB stars in the Galaxy
and argue that the smaller bubbles around several B stars are those
produced by relatively soft radiation which fails to produce
significant HII regions. The nebulae presented here may well represent
such bubbles or could be simple reflection nebulae due to an evolved
generation of B stars.

\section{Discussion}

All the four sample sets of targets used in this study have FIR
luminosities in the range 10$^3$--10$^5$\lsun\, and satisfy similar
FIR colour criteria. Nevertheless, the targets from Fon02 and Faun04
surveys reveal significantly large number of nebulae and redenned
sources in comparision to targets from the Sri02 and Mol96 samples.
To understand these variations among the samples, we plotted the
galactic positions of all targets (Fig.~7) and found that the Fon02
and Faun04 targets are concentrated in the galactic
mid-plane(b=$\pm$2) and close to the galactic center (l=0-40 and
l=280-360) whereas the Sri02 and Mol96 targets are situated relatively
away from the galactic center (l=10-220) and also have a wider
distribution with respect to the mid-plane (b=$\pm$3). The Faun04
sample also contains a large fraction of well known UCHII regions
which are found in the Galactic mid-plane. The HMPO candidates from
the RMS survey also show similar concentration limited to the inner
plane of the galaxy \citep{urquhart07}. Since the Spitzer bands are
well suited to probe the mid-plane and most HMPO candidates are
concentrated in this area, the highest number of nebulae found in the
samples of Fon02 and Faun04 is not a surprise. This result also
indicates that much of the on-going massive star formation is also
found within the Galactic mid-plane in accordance with the known scale
height of 100pc for OB stars. The location of the largest number of
GLIMPSE nebulae in the mid-plane also explains why \citet{kumar06} did
not find clusters in this region where the 2MASS K band suffers high
extinction.

   \begin{figure}
   \centering
   \includegraphics[width=9cm]{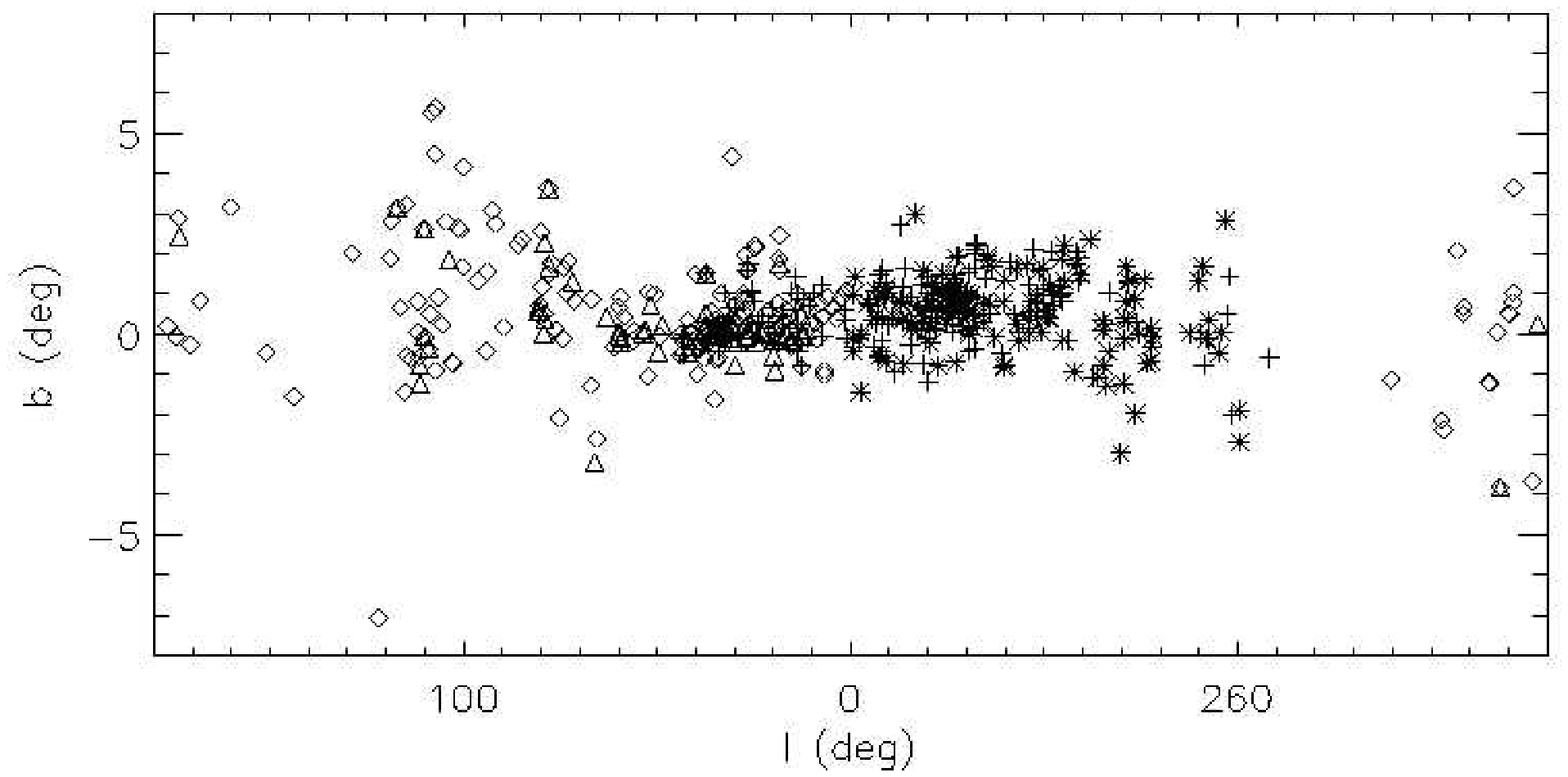}
   
   \caption{Galactic positions of all the targets from \citet{sri02}
   (triangles), \citet{mol96} (diamonds), \citet{fon02} (asterisks)
   and \citet{faun04} (plus marks). }

  \end{figure}

As shown in Sec.\,3 the null result of cluster detection does not
necessarily imply the absence of embedded clusters. For example, deep
near-infrared imaging of the source I19343+2026 (at 0.5\arcsec seeing
and K =19mag limit) shows a rich cluster at 2\mum\, and only a few
stars in the Spitzer data. This may suggest that many of the compact
nebulae found in the Spitzer data may indeed be undetected embedded
clusters. In such an event, the cluster statistics of 25\% derived by
\citet{kumar06} is clearly a lower limit for the HMPO candidates. This
result will have important consequences on the nature and evolutionary
state of the high mass dense cores studied so far.

The point sources detected from the GLIMPSE survey show colors and
magnitudes representative of young stellar objects in the mass range
8\msun--20\msun\, and in the evolutionary stages between Class I and
III. The high $\alpha$ values indicate a steeply rising Spectral
Energy Distribution in the 3.6--8.0\mum\, region and mostly show
non-detections in the 2\mum\, K band from 2MASS. The radiative
transfer models database from RWIWD06 was used to examine the
individual components of emission for each model, namely the star,
disk and envelope contributions. In these models by RWIWD06, radiation
arising from each of the entities namely star, disk and envelope are
computed. For example, the envelope emission in these models are
calculated as radiation originating in the envelope but not
necessarily produced inside the envelope. Reprocessed star light for
which the envelope is the starting point is treated as envelope
emission. Comparing the individual model components from star,disk and
envelope, we find that the observed data in IRAC bands are best
described by the envelope emission component. In most cases, the
stellar photosphere and/or the accreting disk have very little or
absent contribution in the IRAC bands. Therefore the high $\alpha$
values in the IRAC bands is mostly due to emission arising in
envelopes which is due to reprocessed star and/or star$+$disk
emission. In most cases, the stellar photosphere and/or the accreting
disk have no directly contributing emission in the IRAC bands. Thus we
are witnessing the envelopes of massive young stars through these
observations. The model parameters that fit the observed data well
suggests that a majority of the HMPO candidates host massive
protostars in the mass range 8\msun--20\msun\ and beyond. More
rigorous identification of individual HMPOs can be made by
constructing the SED over a longer wavelength range at similar spatial
resolution and modelling them with radiative transfer analysis. In
Paper II we will discuss the results from such an analysis of
individual massive protostars by building the SED with all available
data in the literature.

Thus far, the analysis presented here indicates the presence of HMPOs
identified by their shining envelopes that are embedded in compact
infrared nebulae. Apart from this infrared view, we know apriori from
the original survey papers of these HMPOs that they are dense cores
with high column densities \citep{beu02}, ongoing infall
\citep{fuller05} and outflow activity \citep{beu02b,zhang05}. High
sensitivity observations with the Very Large Array (VLA) show
unresolved centimeter continuum emission associated with many of these
point sources \citep{mol98,carral99,zapata06}, some of which are also
known to be well identified driving sources of massive molecular
outflows \citep{beu04}. For example in the sample of Mol96, we found
IR counterparts for 27 unresolved VLA sources within a 5\arcsec
radius, of which most had a saturated flux at 5.8\mum\, and 8\mum\,
bands. Six sources had data in all bands and satisfied the criteria of
high AM sources and all these six sources coincide with the VLA peaks
to an accuracy of $\le$3\arcsec. In an effort to separate HMPOs based
on centimeter continuum emission \citet{urquhart07} find several MSX
sources with and without radio continuum emission, implying that a
large fraction of the sources indeed display centimeter continuum
emission along with other signposts of HMPOs such as millimeter and
FIR emission. \citet{carral99} found accurately matching compact 3.6cm
continuum sources associated with the dust emission peaks of 12
sources derived from Sri02 and Mol96 samples. Among them, four sources
lie in the galactic plane covered by GLIMPSE data and we find high AM
sources in all the four sources. The complete list of sources that
satisfy the HMPO criteria and their correlation with the observed
centimeter flux will be discussed in Paper II. The centimeter
continuum emission represent free-free emission in ionized gas, and
thought to trace the ultra-compact or hyper-compact HII regions around
some of the HMPOs situated inside the thick shining envelopes
described above.  From the available data, although it is impossible
to rule out the possibility that all of this emission is arising from
jets, it is also improbable that the centimeter emission in all these
sources are originating in jets \citep[e.g][]{zapata06}. If we assume
that a good fraction of the observed centimeter peaks are HII
components and add the result from this work that they are associated
with IR counterparts satisfying the HMPO criteria along with other
sign-posts of on-going accretion such as outflow and disk, we are then
witnessing a combination of ionised and molecular material around
accreting stars. This may suggest that the accretion is on-going and
likely composed of both molecular and ionised components of gas. The
presence of ionised matter may be particularly true when the embedded
star is close to 20\msun\, or more. Ionised accretion flows are well
demonstrated by the study of UCHII region such as G10.6-0.4
\citep{keto02a} and accretion signatures in the Br$\gamma$ line
profiles (which can represent ionised gas) in some massive young stars
\citep{blum04}.  Combining the implications of observations from the
centimeter and millimeter to the infrared thus leads us to a scenario
where the dense cores are hosting precursors to OB type stars, some of
which have even produced a compact ionized region and continue to
accrete, adding more mass to the central star. Such a process may be
of particular importance in producing O stars.  Therefore these
observations are supportive of massive star formation scenarios
through continuuing accretion process with a combination of both
molecular and ionized gas components \citep{keto02,keto03} rather than
rapid accretion of molecular gas alone \citep{mckeetan03}. Rapid
accretion of molecular gas alone may be prominent only in the very
early (and short time scale) class O phase of the HMPOs, observed in
some ``mm-only'' type of objects \citep[e.g.][]{hill06}.

\section{Summary \& Conclusions}

We have used the GLIMPSE point source catalog and the image cutouts
facility to investigate the properties of candidate massive
protostellar objects distributed over the northern and southern
hemispheres of the sky. Data were found for 381 out of 500 examined
targets. The analysis of the point source photometry and images can be
summarised as follows:

\begin{itemize}

\item{The GLIMPSE data could probe in to the HMPO targets in the 
Galactic mid-plane revealing their IR counterparts. No significant
clustering was observed around HMPO targets, however, multiple
components or isolated bright point sources with intrinsic redenning
were found in most cases indicating multiplicity at birth.}

\item{Color-Color diagram analysis of point sources found in the target 
fields and 40 randomly selected control fields clearly demonstrate the
presence of redenned Class I and II type sources lying in the target
fields. The spectral index ($\alpha$) of such point sources computed
using fluxes in the IRAC bands display high $\alpha$ values of 3-5
suggesting their deeply embedded nature in dense cocoons around the
HMPOs. Absolute magnitude vs alpha-magnitude (AM product) diagrams
demonstrate that these point sources occupy zones that are
representative of massive young stars ranging in the mass between
8--20\msun\ or more, implying the presence of  massive protostars.}

\item{A total of 79 point sources could be classified as HMPO 
candidates, and they display a good spatial correlation with the
associated IRAS sources.}

\item{Nearly 60\% of the targets are surrounded by compact infrared 
nebulae, particularly luminous in the 8\mum\, band. These nebulae
display morphologies similar to UCHII regions such as cometary,
core-halo, shell and bipolar shapes. The size distribution of these
nebulae for the Sri02 and Mol96 sample (which are assumed to be
youngest subsets) display a characteristic size scale of 0.1-1~pc,
with a mean value of 0.5~pc, showing that the nebulae are limited to
the boundaries of the dense cores mapped by the millimeter continuum
emission and may be reflecting the underlying physical structure of
these cores.}

\item{The GLIMPSE view suggests that the massive star forming dense 
cores contain precursors to OB stars shining through their thick
envelopes and surrounded by compact infrared nebulae. The observed
correlation of unresolved centimeter continuum emission from VLA and
the GLIMPSE counterparts suggest that some of these HMPOs have
produced an ultra/hypercompact HII region close to the star and
continue to accrete matter. Therefore both ionized and molecular
components of accretion are likely involved in building the most
massive stars such as O stars. The observations thus favour a scenario
of massive star formation through continuuing accretion involving both
molecular and ionised flows rather than rapid accretion of molecular
gas alone.}

\end{itemize}

\begin{acknowledgements}

We thank the referee Tom Megeath and the editor Malcolm Walmsley for
useful suggestions that has improved the presentation of the paper and
also for providing the Orion data for comparision.  MSNK and JMCG are
supported by a research grant POCTI/CFE-AST/55691/2004 and JMCG is
supported by a doctoral fellowship SFRH/BD/21624/2005 approved by FCT
and POCTI, with funds from the European community programme
FEDER. This research has made use of the NASA/ IPAC Infrared Science
Archive, which is operated by the Jet Propulsion Laboratory,
California Institute of Technology, under contract with the National
Aeronautics and Space Administration.

\end{acknowledgements}

\end{document}